# Quantum entanglement inferred by the principle of maximum Tsallis entropy

Sumiyoshi Abe[1] and A. K. Rajagopal[2]

[1]*College of Science and Technology, Nihon University,*
*Funabashi, Chiba 274-8501, Japan*
[2]*Naval Research Laboratory, Washington D.C., 20375-5320, U.S.A.*

The problem of quantum state inference and the concept of quantum entanglement are studied using a non-additive measure in the form of Tsallis' entropy indexed by the positive parameter q. The maximum entropy principle associated with this entropy along with its thermodynamic interpretation are discussed in detail for the Einstein-Podolsky-Rosen pair of two spin-1/2 particles. Given the data on the Bell-Clauser-Horne-Shimony-Holt observable, the analytic expression is given for the inferred quantum entangled state. It is shown that for q greater than unity, indicating the sub-additive feature of the Tsallis entropy, the entangled region is small and enlarges as one goes into the super-additive regime where q is less than unity. It is also shown that quantum entanglement can be quantified by the generalized Kullback-Leibler entropy.

## I. INTRODUCTION

Entanglement is a fundamental concept highlighting non-locality of quantum mechanics. It is well known [1] that the Bell inequalities, which any local hidden variable theories should satisfy, can be violated by entangled states of a composite system described by quantum mechanics. Experimental results [2] suggest that naive local realism à la Einstein, Podolsky, and Rosen [3] may not actually hold. Related issues arising out of the concept of quantum entanglement are quantum cryptography, quantum teleportation, and quantum computation. To give a quantitative measure of this concept from various physical and mathematical viewpoints is one of the main thrusts of current research in this area. The status of the ongoing research can be found for example in a recent report [4].

Quantum mechanical description of composite states is given in terms of a Hermitian, traceclass, positive semi-definite operator, the density matrix, $\hat{\rho}$. This operator replaces the classical probability concepts in addressing the quantum systems in terms of a more general probability amplitude framework. The classical information theory based on the Shannon entropy measure associated with the classical probability has been successful in our understanding of signal processing and communication. Classical entanglement of a bivariate probability distribution is understood in terms of the mutual entropy which is always positive if the variables are correlated and zero if uncorrelated. The quantum counterpart of the entropic measure is the von Neumann entropy defined in terms of the density matrix, which incorporates both pure and mixed states of a system. It is zero in the case when the state is pure, i.e. $\hat{\rho}^2 = \hat{\rho}$, and nonzero in the case when the state is mixed, i.e. $\hat{\rho}^2 < \hat{\rho}$, the latter being the most commonly occurring feature in practice. Quantum entanglement can occur in both cases. There has been ongoing efforts in describing quantitatively the concept of quantum entanglement in terms of the von Neumann entropy and its variants, such as the Kullback-Leibler entropy.

Jaynes [5] has pioneered in developing and applying the maximum entropy principle for inferring the states of the system be it classical or quantum, with given data as constraints. He also pointed out that the Lagrange multipliers needed to specify the constraints lend themselves to a "thermodynamic-like" description of the system. In a recent paper [6], quantum entanglement has been studied in the context of quantum-state inference using the Jaynes maximum entropy principle. In an important special example, it was shown that this method yields *fake* entanglement when only the data on the expectation value of the Bell-Clauser-Horne-Shimony-Holt (Bell-CHSH) observable [7] is used, in that the density matrix so derived produces apparent entanglement. To rectify this, the authors of Ref. [6] suggested that the Jaynes inference scheme be supplemented with a procedure of minimization of entanglement. Quite recently, one of the present authors [8] suggested a possible cure of this problem by augmenting the data with dispersion of the Bell-CHSH observable. It was found that the Jaynes inference scheme with this set of two independent data can correctly yield the maximum entropy states with true entanglement, which are the intelligent states, that is, states that saturate the uncertainty relation between the data on the expectation value and the dispersion. It was also shown how, in this example, one may interpret the Lagrange multipliers appearing therein, as thermodynamic-like parameters associated with entanglement.

Much of this type of work involved "additivity" and "concavity" of the above-mentioned entropies. For a better understanding of mixed states and even states with fractal behavior, another "additive" measure, called the Rényi entropy [9], with a parameter $\alpha$, has often been invoked. This has been used to understand aspects of quantum entanglement and inseparability of mixed states [10-12]. However this entropy is not convex for all values of the parameter $\alpha$, specifically for $\alpha > 1$. There are also theoretical observations [13-15] that the degree of quantum entanglement between remote systems (defined in a certain manner) mat be preserved or decreased but cannot be increased by local operations on each system. This property analogous to the second law of thermodynamics has motivated researchers to investigate thermodynamic-like description of quantum entangled states [4, 16-19]. A point that seems to stand out from the works [18,19] is that *the measure of quantum entanglement may not be additive*.

A generalization of the Jaynes maximum entropy principle along with an attendant "thermodynamics" to "non-additive" cases has been proposed and widely discussed in the area of non-extensive statistical mechanics. This scheme is based on the Tsallis entropy [20]:

$$S_q[\hat{\rho}] = \frac{1}{1-q}\left(\mathrm{Tr}\,\hat{\rho}^q - 1\right), \tag{1}$$

where $q$ is greater than zero and describes the degree of non-additivity. In the limit $q \to 1$, this entropy converges on the von Neumann entropy: $S[\hat{\rho}] = -\mathrm{Tr}(\hat{\rho}\ln\hat{\rho})$. The Tsallis entropy shares many of the common properties of the von Neumann entropy in that it is concave for all values of $q$, unlike the Rényi entropy, obeys the *H*-theorem, etc. The additivity is modified, however. For the product state $\hat{\rho}^{(1)} \otimes \hat{\rho}^{(2)}$, $S_q$ satisfies the following relation:

$$S_q[\hat{\rho}^{(1)} \otimes \rho^{(2)}] = S_q[\hat{\rho}^{(1)}] + S_q[\rho^{(2)}] + (1-q)S_q[\hat{\rho}^{(1)}]S_q[\rho^{(2)}]. \tag{2}$$

The cases $0 < q < 1$ and $q > 1$ are said to be super-additive and sub-additive, respectively. The corresponding generalized maximum entropy principle is based on the data given in the form constraints defined as normalized *q*-expectation values [21]

$$<\hat{B}>_q = \frac{\mathrm{Tr}(\hat{\rho}^q \hat{B})}{\mathrm{Tr}(\hat{\rho}^q)} \tag{3}$$

and the normalization condition $\mathrm{Tr}\,\hat{\rho} = 1$. The resulting maximum entropy states are found to be never factorized even if the operator $\hat{B}$ is the sum of two mutually commuting observables and consequently correlation is always induced [22]. Accordingly, we may surmise that this correlation is linked with quantum entanglement produced by the long arm of inseparability.

The purpose of this paper is therefore to reexamine the example studied in Refs. [6,8] within the framework of the principle of maximum Tsallis entropy with two constraints as in Ref. [8]. We suggest here that the Tsallis entropy is a valid choice for this purpose by virtue of its concavity for all values of the parameter $q$. Also, unlike the Rényi entropy, it has built into it the feature that even if the density matrix is separable, this entropy manifests non-additivity of the partial entropies of the individual components, while the normalized *q*-expectation values defined in eq. (3) remain additive. It seems that the long arm of inseparability may be another feature of non-additivity and thus lends itself, naturally as an information-theoretic quantity, to investigate the property of entanglement mentioned above. Moreover, the maximum entropy principle holds for the Tsallis entropy with attendant thermodynamic consequences for describing non-additive features. This allows us to examine if indeed one can associate a thermodynamic-like description of entangled states using the Tsallis maximum entropy ideas in this context. Since this scheme has the features of non-additivity of the entropy but additivity of the normalized *q*-expectation values [21], it seems to fit a thermodynamic-like description of entanglement. Thus it is appropriate to examine this aspect of the theory in the present context.

This paper is organized as follows. In Sec. II, we apply the Tsallis maximum entropy principle to statistical inference of entangled quantum states. An analytic solution of this procedure is given. In Sec. III, we consider the generalized Kullback-Leibler entropy as the non-additive measure of quantum entanglement and explicitly calculate it for the inferred state obtained in Sec. II. In Sec. IV, we develop generalized thermodynamics of the inferred state. Section V is devoted to concluding remarks.

## II. QUANTUM ENTANGLEMENT OF MAXIMUM TSALLIS ENTROPY STATES

As in Refs. [6,8], we consider the Einstein-Podolsky-Rosen pair of two spin-1/2 particles, $A$ and $B$, and take the Bell-CHSH observable [7], which is given in terms of the Pauli matrices by

$$\hat{B} = \sqrt{2}\left(\sigma_{Ax} \otimes \sigma_{Bx} + \sigma_{Az} \otimes \sigma_{Bz}\right). \tag{4}$$

In the Bell basis [23]

$$\left|\Phi^{\pm}\right\rangle = \frac{1}{\sqrt{2}}\left(\left|\uparrow\right\rangle_A\left|\uparrow\right\rangle_B \pm \left|\downarrow\right\rangle_A\left|\downarrow\right\rangle_B\right), \quad \left|\Psi^{\pm}\right\rangle = \frac{1}{\sqrt{2}}\left(\left|\uparrow\right\rangle_A\left|\downarrow\right\rangle_B \pm \left|\downarrow\right\rangle_A\left|\uparrow\right\rangle_B\right), \tag{5}$$

this operator is expressed as follows:

$$\hat{B} = 2\sqrt{2}\left(\left|\Phi^{+}\right\rangle\left\langle\Phi^{+}\right| - \left|\Psi^{-}\right\rangle\left\langle\Psi^{-}\right|\right). \tag{6}$$

Its $q$-expectation value

$$<\hat{B}>_q = b_q \tag{7}$$

lies in the range

$$0 \leq b_q \leq 2\sqrt{2}. \tag{8}$$

To apply the principle of maximum Tsallis entropy to statistical inference of quantum states, it is necessary to specify a set of the statistical data. As in Ref. [8], in addition to the expectation value, we employ its dispersion

$$<\hat{B}^2>_q = \sigma_q^2, \tag{9}$$

which satisfies

$$\sigma_q^2 \leq 8. \tag{10}$$

The reasons why we include this are that, firstly

$$\hat{B}^2 = 8\left(|\Phi^+\rangle\langle\Phi^+| + |\Psi^-\rangle\langle\Psi^-|\right) \tag{11}$$

is linearly independent of $\hat{B}$, secondly this operator obviously commutes with $\hat{B}$, and, thirdly any measured data on $\hat{B}$ have to be accompanied with its reliability, a measure of which is given by the dispersion. Clearly, $\sigma_q^2 \geq b_q^2$. On the other hand, the Schwarz inequality, $<\hat{X}^2>_q <\hat{Y}^2>_q \geq <\hat{X}\hat{Y}>_q^2$, with the identification $\hat{X} = \hat{B}$, $\hat{Y} = \hat{B}^2$ and the observation $\hat{X}^2 = \hat{B}^2$, $\hat{Y}^2 = 8\hat{B}^2$, $\hat{X}\hat{Y} = 8\hat{B}$, leads to the uncertainty relation

$$\sigma_q^2 \geq 2\sqrt{2}\, b_q. \tag{12}$$

Now, under the two constraints on the data in eqs. (7) and (9) together with the normalization condition, maximization of the Tsallis entropy in eq. (1) yields the following optimal state

$$\hat{\rho}_{qAB}(\lambda_1, \lambda_2) = \frac{1}{Z_q(\lambda_1, \lambda_2)}$$
$$\times \left\{\left[1 + \frac{1-q}{c_q}\left(\lambda_1 b_q + \lambda_2 \sigma_a^2\right)\right]\hat{I}_{AB} - \frac{1-q}{c_q}\left(\lambda_1 \hat{B} + \lambda_2 \hat{B}^2\right)\right\}^{1/(1-q)}, \tag{13}$$

$$Z_q(\lambda_1, \lambda_2) = \mathrm{Tr}\left\{\left[1 + \frac{1-q}{c_q}\left(\lambda_1 b_q + \lambda_2 \sigma_a^2\right)\right]\hat{I}_{AB} - \frac{1-q}{c_q}\left(\lambda_1 \hat{B} + \lambda_2 \hat{B}^2\right)\right\}^{1/(1-q)}, \tag{14}$$

where $\hat{I}_{AB}$ is the unit operator of the composite $AB$-system and $\lambda_1$ and $\lambda_2$ are the Lagrange multipliers associated with the constraints on the data and the quantity $c_q$ is defined by

$$c_q = \mathrm{Tr}\left(\hat{\rho}_{qAB}\right)^q. \tag{15}$$

From eqs. (13)-(15), follows a general relation

$$c_q = \left[Z_q(\lambda_1, \lambda_2)\right]^{1-q}. \tag{16}$$

In spite of the fact that the operator appearing in the curly brackets in eq. (13) is nothing but the sum of two mutually commuting operators $|\Phi^+\rangle\langle\Phi^+|$ and $|\Psi^-\rangle\langle\Psi^-|$, the density matrix is never factorizable due to its non-exponential form. Thus, correlation is induced by non-additivity of the formalism. Such correlation clearly disappears when $q \to 1$, since in this limit equations (13) and (14) acquire the ordinary exponential form obtained from the standard Jaynes scheme.

To manipulate the density matrix in eq. (13), we use eqs. (6) and (11) as well as the

completeness relation

$$|\Phi^+\rangle\langle\Phi^+| + |\Phi^-\rangle\langle\Phi^-| + |\Psi^+\rangle\langle\Psi^+| + |\Psi^-\rangle\langle\Psi^-| = \hat{I}_{AB}, \qquad (17)$$

and express it in the form

$$\hat{\rho}_{qAB} = \frac{1}{Z_q(\lambda_1, \lambda_2)} \times \left\{ \mu_{q0}\left(|\Phi^-\rangle\langle\Phi^-| + |\Psi^+\rangle\langle\Psi^+|\right) + \mu_{q-}|\Phi^+\rangle\langle\Phi^+| + \mu_{q+}|\Psi^-\rangle\langle\Psi^-| \right\}^{1/(1-q)}, \qquad (18)$$

where

$$\mu_{q0} = 1 + \frac{1-q}{c_q}\left(\lambda_1 b_q + \lambda_2 \sigma_q^2\right), \qquad (19)$$

$$\mu_{q\pm} = 1 + \frac{1-q}{c_q}\left[\lambda_1\left(b_q \pm 2\sqrt{2}\right) + \lambda_2\left(\sigma_q^2 - 8\right)\right]. \qquad (20)$$

Since $|\Phi^+\rangle\langle\Phi^+|$, $|\Phi^-\rangle\langle\Phi^-|$, $|\Psi^+\rangle\langle\Psi^+|$, and $|\Psi^-\rangle\langle\Psi^-|$ are mutually orthogonal projection operators, the above density matrix is found to be

$$\hat{\rho}_{qAB} = \frac{1}{Z_q(\lambda_1, \lambda_2)} \left[ \left(\mu_{q0}\right)^{1/(1-q)}\left(|\Phi^-\rangle\langle\Phi^-| + |\Psi^+\rangle\langle\Psi^+|\right) + \left(\mu_{q-}\right)^{1/(1-q)}|\Phi^+\rangle\langle\Phi^+| + \left(\mu_{q+}\right)^{1/(1-q)}|\Psi^-\rangle\langle\Psi^-| \right], \qquad (21)$$

and accordingly

$$Z_q(\lambda_1, \lambda_2) = 2\left(\mu_{q0}\right)^{1/(1-q)} + \left(\mu_{q-}\right)^{1/(1-q)} + \left(\mu_{q+}\right)^{1/(1-q)}. \qquad (22)$$

By the same reason, the factor $c_q = \mathrm{tr}(\hat{\rho}_{qAB}^q)$ is also calculated to be

$$c_q = \frac{1}{[Z_q(\lambda_1, \lambda_2)]^q}\left[2\left(\mu_{q0}\right)^{q/(1-q)} + \left(\mu_{q-}\right)^{q/(1-q)} + \left(\mu_{q+}\right)^{q/(1-q)}\right]. \qquad (23)$$

Then, from the identical relation in eq. (16), we find another expression for $Z_q$:

$$Z_q(\lambda_1, \lambda_2) = 2\left(\mu_{q0}\right)^{q/(1-q)} + \left(\mu_{q-}\right)^{q/(1-q)} + \left(\mu_{q+}\right)^{q/(1-q)}. \qquad (24)$$

$b_q = <\hat{B}>_q$ and $\sigma_q^2 = <\hat{B}^2>_q$ can also be calculated in a similar way:

$$b_q = \frac{2\sqrt{2}}{Z_q(\lambda_1, \lambda_2)}\left[\mu_{q-}^{q/(1-q)} - \mu_{q+}^{q/(1-q)}\right], \qquad (25)$$

$$\sigma_q^2 = \frac{8}{Z_q(\lambda_1, \lambda_2)} \left[ \mu_{q-}^{q/(1-q)} + \mu_{q+}^{q/(1-q)} \right]. \tag{26}$$

We wish to express the elements of the density matrix in terms of the values of the data $b_q$ and $\sigma_q^2$. For this purpose, some algebraic manipulations are needed. From eq. (20), the Lagrange multipliers are found to be

$$\lambda_1 = \frac{c_q}{4\sqrt{2}(1-q)} (\mu_{q+} - \mu_{q-}), \tag{27}$$

$$\lambda_2 = \frac{c_q}{(\sigma_q^2 - 8)(1-q)} \left\{ \frac{1}{2}\left(1 - \frac{b_q}{2\sqrt{2}}\right)\mu_{q+} + \frac{1}{2}\left(1 + \frac{b_q}{2\sqrt{2}}\right)\mu_{q-} - 1 \right\}. \tag{28}$$

Also, from eqs. (25) and (26), it follows that

$$(\mu_{q\pm})^{1/(1-q)} = \left[ \frac{1}{16}(\sigma_q^2 \mp 2\sqrt{2}\, b_q) Z_q(\lambda_1, \lambda_2) \right]^{1/q}. \tag{29}$$

Using eqs. (27)-(29) in eq. (19), we have

$$(\mu_{q0})^{1/(1-q)} = \left[ \frac{1}{16}(8 - \sigma_q^2) Z_q(\lambda_1, \lambda_2) \right]^{1/q}. \tag{30}$$

Substitution of eqs. (29) and (30) into eq. (22) leads to

$$(Z_q)^{(q-1)/q} = 2\left( \frac{8 - \sigma_q^2}{16} \right)^{1/q} + \left( \frac{\sigma_q^2 - 2\sqrt{2}\, b_q}{16} \right)^{1/q} + \left( \frac{\sigma_q^2 + 2\sqrt{2}\, b_q}{16} \right)^{1/q}. \tag{31}$$

Therefore we find the following closed expression for the density matrix:

$$\hat{\rho}_{qAB} = \frac{1}{(Z_q)^{(q-1)/q}} \left\{ \left( \frac{8 - \sigma_q^2}{16} \right)^{1/q} (|\Phi^-\rangle\langle\Phi^-| + |\Psi^+\rangle\langle\Psi^+|) \right.$$
$$\left. + \left( \frac{\sigma_q^2 + 2\sqrt{2}\, b_q}{16} \right)^{1/q} |\Phi^+\rangle\langle\Phi^+| + \left( \frac{\sigma_q^2 - 2\sqrt{2}\, b_q}{16} \right)^{1/q} |\Psi^-\rangle\langle\Psi^-| \right\}, \tag{32}$$

where $(Z_q)^{(q-1)/q}$ is given in eq. (31). In the limit $q \to 1$, this becomes the density matrix given in Ref. [8]. On the other hand, in the limit $q \to \infty$, $\hat{\rho}_{qAB}$ becomes reduced to $(1/4)\hat{I}_{AB}$. This is a product state since $\hat{I}_{AB} = \hat{I}_A \otimes \hat{I}_B$, where $\hat{I}_A$ ($\hat{I}_B$) is the unit operator in the space of the spin $A$ ($B$).

The separability condition given in Ref. [12] states that the eigenvalues of the density matrix do not exceed $1/2$. That is, it is sufficient for realizing entanglement if the largest eigenvalue obeys the condition

$$\left[ \frac{\sigma_q^2 + 2\sqrt{2}\, b_q}{16(Z_q)^{q-1}} \right]^{1/q} > \frac{1}{2}. \tag{33}$$

This equation tells how quantum entanglement is induced simultaneously by the maximum entropy principle and non-additivity of the Tsallis entropy. In Fig. 1, the condition in eq. (33) is depicted for some values of $q$. The entangled region is shown as the "island" above the "sea level" $z = 1/2$. It is interesting to observe that $q > 1$, indicating the sub-additive feature of the Tsallis entropy, the entangled region is small and enlarges as we go into the super-additive regime where $q < 1$.

Closing this section we wish to note that the intelligent state, which saturates the uncertainty relation in eq. (12), does not have the element of $|\Psi^-\rangle\langle\Psi^-|$. Moreover, if $b_q = 2\sqrt{2}$ (i.e., $\sigma_q^2 = 8$), then we see the purification of the state: $\hat{\rho}_{qAB}^{\text{pure}} = |\Phi^+\rangle\langle\Phi^+|$, which is a maximally entangled state. This point will be revisited in Sec. IV from the thermodynamic viewpoint.

### III. GENERALIZED KULLBACK-LEIBLER MEASURE OF QUANTUM ENTANGLEMENT

To quantitatively describe the concept of quantum entanglement, the entropic measures such as the von Neumann, Kullback-Leibler, and Rényi entropies have been investigated in the literature [11-15]. These measures are additive under replication (or, multi-plication) of the system. As discussed in Refs. [18,19], however, a measure of quantum entanglement may not be additive, in general. What is more essential is the convergence of the "thermodynamic limit", in which the system degrees of freedom become arbitrarily large. The non-additive approach we have developed in the preceding section is based on non-extensive Tsallis statistical mechanics, which is known to lead to a consistent thermodynamic framework. In this section, we consider as a non-additive measure of quantum entanglement the generalized Kullback-Leibler (KL) entropy associated with the Tsallis entropy and calculate its value for the inferred entangled state given in Sec. II.

The generalized KL entropy of $\hat{\rho}$ with respect to the reference state $\hat{\rho}'$ is defined by

$$K_{q'}[\hat{\rho}, \hat{\rho}'] = \frac{1}{1-q'} \text{Tr}\left[ \hat{\rho}^{q'} \left( \hat{\rho}^{1-q'} - \hat{\rho}'^{1-q'} \right) \right], \tag{34}$$

which is the quantum mechanical generalization of the classical definition given in Refs. [24]. (See also Ref. [25].) This quantity is positive semi-definite and is equal to zero if and only if $\hat{\rho} = \hat{\rho}'$. It measures the difference between $\hat{\rho}$ and $\hat{\rho}'$. The parameter $q'$ is taken differently from $q$ of the Tsallis entropy for the sake of generality. In the limit $q' \to 1$, the generalized KL entropy becomes the ordinary KL entropy: $K_{q'}[\hat{\rho}, \hat{\rho}'] \to \text{Tr}[\hat{\rho}(\ln \hat{\rho} - \ln \hat{\rho}')]$.

Consider the marginal density matrices of the inferred density matrix $\hat{\rho}_{qAB}$, which are defined as follows:

$$\hat{\rho}_{qA} = \text{Tr}_B \hat{\rho}_{qAB}, \qquad \hat{\rho}_{qB} = \text{Tr}_A \hat{\rho}_{qAB}, \qquad (35)$$

where the symbol $\text{Tr}_A$ ($\text{Tr}_B$) stands for the partial trace over the states of the spin $A$ ($B$). The state is unentangled if the total density matrix is the product of the marginal density matrices. Therefore the generalized KL entropy $K_{q'}[\hat{\rho}_{qAB}, \hat{\rho}_{qA} \otimes \hat{\rho}_{qB}]$, which is referred to as the generalized mutual entropy, measures the degree of entanglement in the state $\hat{\rho}_{qAB}$. Note that it is invariant under the local unitary transformation: $\hat{\rho}_{qAB} \to \hat{U}_A \otimes \hat{U}_B \hat{\rho}_{qAB} \hat{U}_A^\dagger \otimes \hat{U}_B^\dagger$. (Though the generalized conditional entropies might also be considered for discussing the information content given the marginal density matrices, the generalized mutual entropy considered here seems to offer the most suitable measure.) In the special example considered here, we have

$$\text{Tr}_A |\Phi^\pm\rangle\langle\Phi^\pm| = \text{Tr}_A |\Psi^\pm\rangle\langle\Psi^\pm| = \frac{1}{2} \hat{I}_B, \qquad (36)$$

$$\text{Tr}_B |\Phi^\pm\rangle\langle\Phi^\pm| = \text{Tr}_B |\Psi^\pm\rangle\langle\Psi^\pm| = \frac{1}{2} \hat{I}_A. \qquad (37)$$

Using these equations in eq. (32), we find the following marginal density matrices:

$$\hat{\rho}_{qA} = \frac{1}{2} \hat{I}_A, \qquad (38)$$

$$\hat{\rho}_{qB} = \frac{1}{2} \hat{I}_B. \qquad (39)$$

Notably, these are independent of both the data on the Bell-CHSH observable and the parameter $q$. Thus we obtain

$$\hat{\rho}_{qA} \otimes \hat{\rho}_{qB} = \frac{1}{4} \hat{I}_{AB}. \qquad (40)$$

Now, the generalized mutual entropy can immediately be calculated. The result is

$$K_{q'}[\hat{\rho}_{qAB}, \hat{\rho}_{qA} \otimes \hat{\rho}_{qB}]$$

$$= \frac{1}{1-q'} \left\{ 1 - \frac{(Z_q)^{q'(1-q)/q}}{2^{2(1-q')}} \left[ 2\left(\frac{8-\sigma_q^2}{16}\right)^{q'/q} + \left(\frac{\sigma_q^2 - 2\sqrt{2} b_q}{16}\right)^{q'/q} + \left(\frac{\sigma_q^2 + 2\sqrt{2} b_q}{16}\right)^{q'/q} \right] \right\},$$

(41)

where $Z_q$ is given in eq. (31). In the particular case of the intelligent pure state $\hat{\rho}_{qAB}^{\text{pure}} = |\Phi^+\rangle\langle\Phi^+|$ corresponding to $\sigma_q^2 = 2\sqrt{2}\, b_q$ and $b_q = 2\sqrt{2}$, equation (41) is simplified to

$$K_{q'}\left[\hat{\rho}_{qAB}^{\text{pure}},\ \hat{\rho}_{qA} \otimes \hat{\rho}_{qB}\right] = \frac{1}{1-q'}\left[1 - \frac{1}{2^{2(1-q')}}\right], \qquad (42)$$

which is independent of $q$ and, in the limit $q' \to 1$, converges on the value $2\ln 2$. We can find from eq. (41) that quantum entanglement in the super-additive and sub-additive regimes is quantified by the positive semi-definite $K_{q'}$, in conjunction with the results exhibited in Fig. 1.

## IV. GENERALIZED THERMODYNAMICS OF THE INFERRED STATE

The discussion in Sec. II is suitable for thermodynamic consideration of the inferred quantum entangled state since it is developed in conformity with non-extensive Tsallis statistical mechanics. In this section, we present the thermodynamic Legendre transform structure and some equilibrium thermodynamic quantities constructed from the inferred state.

First of all, let us see that the state in eq. (32) is in fact the maximum Tsallis entropy state. The Tsallis entropy of this state is given by

$$S_q[\hat{\rho}_{qAB}] = \frac{1}{1-q}\left\{\text{Tr}[\hat{\rho}_{qAB}(\lambda_1, \lambda_2)]^q - 1\right\}$$

$$= \frac{1}{1-q}\left\{[Z_q(\lambda_1, \lambda_2)]^{1-q} - 1\right\}, \qquad (43)$$

where the identical relation in eq. (16) has been used. Taking the derivatives of this entropy with respect to the Lagrange multipliers $\lambda_1$ and $\lambda_2$ and using eqs. (19), (20), (23)-(26), we can explicitly ascertain the maximum Tsallis entropy condition

$$\left.\frac{\partial S_q[\hat{\rho}_{qAB}]}{\partial \lambda_1}\right|_{b_q, \sigma_q^2} = \left.\frac{\partial S_q[\hat{\rho}_{qAB}]}{\partial \lambda_2}\right|_{b_q, \sigma_q^2} = 0. \qquad (44)$$

We now note that the following relations hold for the Lagrange multipliers:

$$\left.\frac{\partial S_q[\hat{\rho}_{qAB}]}{\partial b_q}\right|_{\lambda_1,\lambda_2} = \lambda_1, \qquad \left.\frac{\partial S_q[\hat{\rho}_{qAB}]}{\partial \sigma_q^2}\right|_{\lambda_1,\lambda_2} = \lambda_2. \qquad (45)$$

These relations allow us to formally define the "free energy"

$$F_q = \lambda_1 b_q + \lambda_2 \sigma_q^2 - S_q, \qquad (46)$$

where

$$S_q = \frac{1}{1-q}\left[(Z_q)^{1-q} - 1\right]. \qquad (47)$$

From this, we obtain

$$\left.\frac{\partial F_q}{\partial \lambda_1}\right|_{b_q,\sigma_q^2} = b_q, \qquad \left.\frac{\partial F_q}{\partial \lambda_2}\right|_{b_q,\sigma_q^2} = \sigma_q^2, \qquad (48)$$

conversely. These steps establish the Legendre transform structure promised earlier.

It is also of interest to examine the limits $\lambda_1, \lambda_2 \to \infty$. These limits essentially correspond to the "zero temperature limit", in which the Tsallis entropy in eq. (47) should vanish, that is, $Z_q \to 1$. Now, from eq. (31), we see that this condition is equivalent to

$$b_q \to 2\sqrt{2}, \qquad \sigma_q^2 \to 8, \qquad (49)$$

which lead to, as expected, the purification of the state: $\hat{\rho}_{qAB} \to |\Phi^+\rangle\langle\Phi^+|$.

## V. CONCLUDING REMARKS

We have studied in an example of two correlated spins the problem of quantum state inference using a non-additive measure in the form of Tsallis' entropy and have presented the analytic solution for the density matrix. By analyzing the condition of entanglement and the positive semi-definiteness of the generalized Kullback-Leibler entropy, the super- and sub-additive regimes of quantum entanglement have been elucidated. We have thus shown how quantum entanglement is induced by the principle of maximum Tsallis entropy with the data on the Bell-CHSH observable and have quantified the degree of entanglement by means of the generalized Kullback-Leibler entropy. We have found that strong non-additivity enhances entanglement in the super-additive case. We have also given a thermodynamic interpretation to the inferred quantum entangled state.

The present approach based on non-extensive Tsallis statistical mechanics has an attractive feature. The generalized "internal energy" defined by the normalized $q$-expectation value of the Bell-CHSH observable regarded as the "Hamiltonian" satisfies additivity [21], whereas the generalized "free energy" is non-additive. This feature has a parallelism with an intriguing analogy between measures of entanglement and thermodynamic quantities pointed out in Ref. [19].

## ACKNOWLEDGMENTS

S. A. acknowledges support from Atomic Energy Research Institute of Nihon University. A. K. R. is supported in part by the Office of Naval Research.

# Figure caption

Fig. 1 The plots of the condition in eq.(33) in the domain $\{0 \leq b_q \leq 2\sqrt{2}, \sigma_q^2 \leq 8\}$: The plane $z=1/2$ and $z = \left[ \left( \sigma_q^2 + 2\sqrt{2}\, b_q \right) \big/ 16 (Z_q)^{q-1} \right]^{1/q}$ with $\sigma_q^2 \geq 2\sqrt{2}\, b_q$ for (a) q=5, (b) q=2, (c) q=1.5, (d) q=0.9, (e) q=0.5, (f) q=0.1.

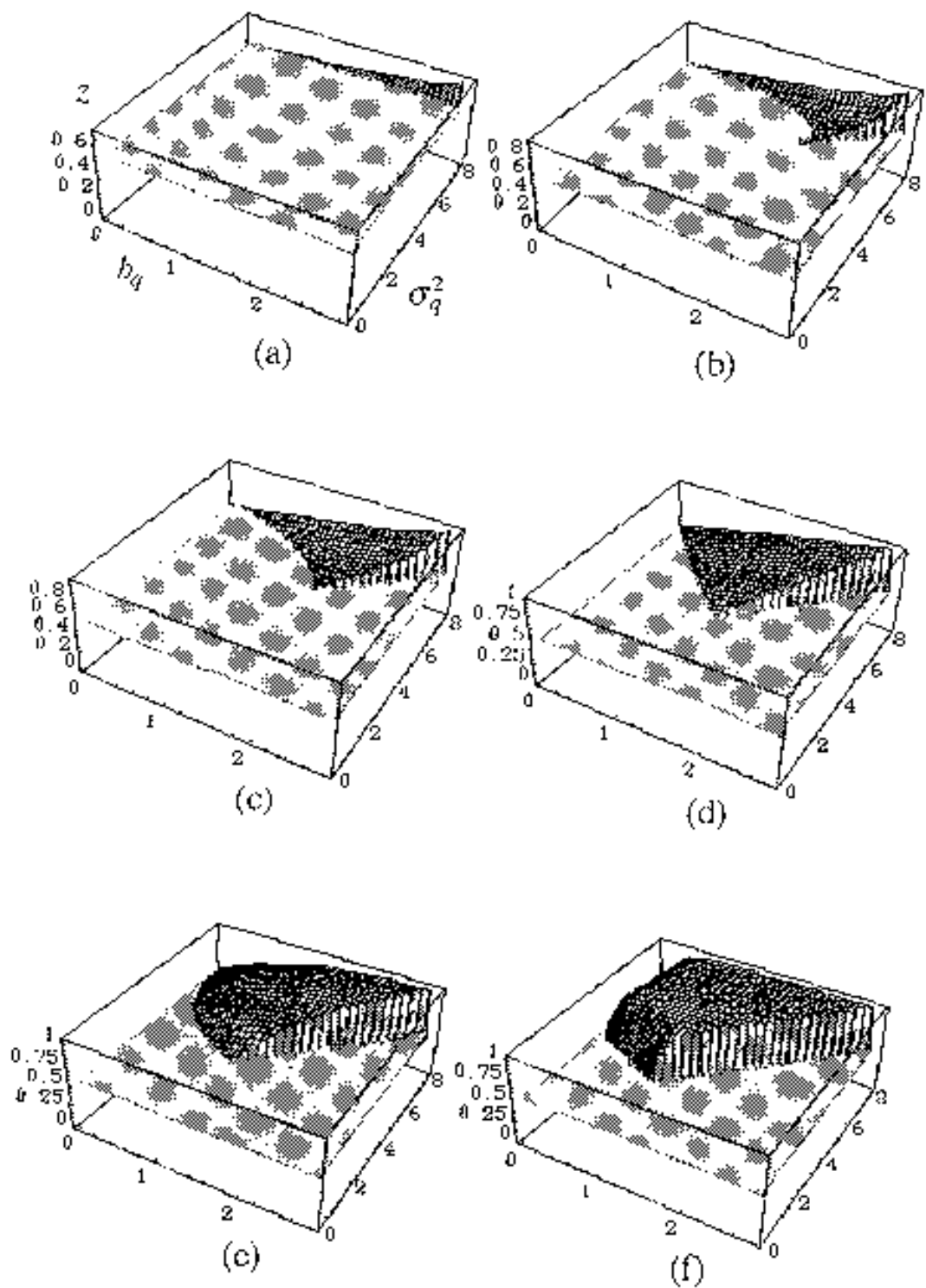

Fig. 1